\documentclass[11pt paper]{article}
\usepackage{amsmath}
\usepackage{graphicx}
\usepackage{setspace}
\usepackage{calc}
\usepackage{color}
\newcommand{\e}{\text{E}\,}
\newcommand{\eq}[1]{Eq.\,\eqref{#1}}

\title{An asymmetric ARCH model and the non-stationarity of Clustering and Leverage effects}

\author{Xin Li\thanks{School of Physics and Astronomy, University of Minnesota, Minneapolis, MN 55455.}\, and Carlos F. Tolmasky\thanks{Institute for Mathematics and Its Applications and MCFAM, University of Minnesota, Minneapolis, MN 55455.}}

\date{}

\begin{document}

\maketitle

\begin{abstract}
We propose a new volatility model based on two stylized facts of the volatility in the stock market: clustering and leverage effect. We calibrate our model parameters, in the leading order, with $77$ years Dow Jones Industrial Average data. We find in the short time scale ($10$ to $50$ days) the future volatility is sensitive to the sign of past returns, i.e. asymmetric feedback or leverage effect. However, in the long time scale ($300$ to $1000$ days) clustering becomes the main factor. We study non-stationary features by using moving windows and find that clustering and leverage effects display time evolutions that are rather nontrivial. The structure of our model allows us to shed light on a few surprising facts recently found by Chicheportiche and Bouchaud \cite{qrBouchaud1}.
\end{abstract}

\section{Introduction}

For the past $40$ years there have been various efforts to model the processes that govern the volatility of financial assets. The interest comes, to a large extent, from the role that the volatility parameter plays in the Black and Scholes' theory. Even when in its original form Black-Scholes' model fails to account for relevant observed effects, such as the smile, the study of the evolution of the volatility process has only gained relevance over time. The reason for this is that doing away with the assumption that the parameter is constant and adding a stochastic component to it helps in the understanding of the smile effect. In all fairness, it is hard to make the case for a constant volatility of asset prices.

A crucial step towards modeling volatilities is the identification of stylized facts that we would like our model to accommodate. Two of the most important stylized facts are clustering and the leverage effect. Clustering refers to the fact that volatility tends to persist: high (low) volatility periods tend to be followed by high (low) volatility periods. The leverage effect, on the other hand, refers to the asymmetry with which volatility moves in reaction to positive returns as opposed to negative returns.

Among the models that try to deal with the volatility by adding stochasticity we can differentiate between two different types:  the ARCH-GARCH family of models and the stochastic volatility models. The former follows the ideas that originated in the time series literature. The autoregressive conditional heteroskedastic (ARCH) model was introduced by Engle \cite{Engle} in $1982$ and a lot of modifications of it have been studied ever since, examples of those modifications are the GARCH$(p,q)$, IGARCH (integrated), EGARCH (exponential), QARCH (quadratic) and TGARCH (threshold) models. Even in both ARCH and GARCH models the return effects are in a symmetric fashion, some of their refinements mentioned building in the capability of accounting for the leverage effect.

In a recent paper, Chicheportiche and Bouchaud \cite{qrBouchaud1} present a thorough study of the quadratic ARCH model originally due to Sentana \cite{quadraticARCH}. Its most general form reads
\begin{equation}
\sigma_t^2 = s^2+\sum_{\tau = 1}^{\infty} \mathcal{L}(\tau) r(t-\tau) + \sum_{\tau,\tau'=1}^{\infty} \mathcal{K}(\tau,\tau') r(t-\tau) r(t-\tau')
\label{qr1}
\end{equation}
where $r(t)$ is the return defined as $r(t) = \text{ln}p(t+1)-\text{ln}p(t)$, $p(t)$ is the price of an asset at time $t$. $\sigma_t^2$ is the conditional variance of $r(t)$. $\mathcal{L}(\tau)$ and $\mathcal{K}(\tau,\tau')$ are some kernels that characterize how the past returns affect the conditional variance. By calibrating the model with US stock returns, Chicheportiche and Bouchaud found that the feedback kernels exhibit some unexpected features. One of their central results is that the off-diagonal terms of the quadratic kernel, namely $\mathcal{K}(\tau,\tau')$ with $\tau \neq \tau'$, while being significant, is one order of magnitude smaller than the diagonal elements $\mathcal{K}(\tau,\tau)$. In addition, the off-diagonal terms are found to be negative for large lags. They also found $\mathcal{K}(\tau,\tau)$ decays as a power-law of the lag, in the line with the long-memory of volatility, and $\mathcal{L}(\tau)$ has two time scales as shown by their two-exponential fit. They argued that the observed complex structure in the kernels is surprising and incompatible with the models proposed so far. We will see later that, indeed, our model may provide a simple interpretation of those features. 

In ARCH-GARCH family of models, the central object to model is the conditional volatility $\sigma_t$. After the dynamics of $\sigma_t$ is specified such as \eq{qr1}, the instantaneous return is constructed as $r(t) = \sigma_t \epsilon(t)$, where $\epsilon(t)$ is independently drawn from a prescribed distribution with zero mean and variance equal to one (e.g.,standard normal distribution). The dynamic of $\sigma_t$ has been specified by a large variety of models in many different ways. A common way is to model how conditional variance $\sigma_t^2$ depends on the quantities like $r(t-\tau)$ or $\sigma_{t-\tau}$ in the past. However, people also model conditional volatility $\sigma_t$ directly or more generally $\sigma_t^\delta$. For a review of those models and how they are compared to the simple GARCH(1,1), we refer the interested readers to the paper by Hansen and Lunde \cite{HansenGARCH}. In this paper, a slightly different approach is adopted to model financial time series. We start with a general assumption that the $\sigma_t$ depends on the past returns $\{r(t-\tau),\tau>0\}$, which it is the assumption that all ARCH-like models are based on, then assume that the feedback from the past returns to the current volatility is ``small" and work in the framework of perturbation theory. We seek the simplest model in our framework that can capture the stylized facts of the data, namely clustered volatility and leverage effect. The specific model we found turns out to be very similar to TGARCH model studied by Zakoian\cite{Zakoian}, however as shown later, instead of performing careful statistical tests of models as most studies on TGARCH model did, we focus on looking for qualitative features that model parameters exhibit when the model is calibrated with data. In this way, we hope our analysis may provide some useful insights about the stock market.  

The rest of paper is organized as follows. Section 2 discusses the general assumptions and guidelines that lead to our asymmetric volatility model. Its relation to other models is briefly mentioned. Section 3 defines various correlation functions and their leading order expressions are calculated within our model. Section 4 performs an empirical study under the stationary assumption, i.e., assuming the model parameters are constant during the whole period. The features of the model parameters are discussed. Section 5 relaxes the stationary assumption to show how those parameters change with time and the nontrivial patterns in time evolution are found. Finally, Section 6 summarizes our findings and concludes the paper. 

\section{The Asymmetric ARCH Model}
We propose in this section the asymmetric volatility model based on some general assumptions. The return of a stock at time $t$ is defined as usual $r(t) = \text{ln} (p(t+1)/ p(t))$, where $p(t)$ is the price of a stock. The $r(t)$ is a random variable and can be written in a form of $r(t) = \sigma(t) \epsilon(t)$. We regard both $\sigma(t)$ and $\epsilon(t)$ are random. $\epsilon(t)$ is independent of the past with the conditions $\e \epsilon(t) = 0$ and $\e \epsilon^2(t) = 1$. $\e$ denotes the unconditional expectation value. If the $\sigma(t)$ is deterministic after conditioning on all past information up to $t$, then the model is much like ARCH-GARCH models; If $\sigma(t)$ is still random even when all past information is given, then the model belongs to a stochastic volatility model. We further assume the unconditional expectation $\e \sigma(t) = \sigma_0(t)$, where $\sigma_0(t)$ is a deterministic function of time $t$, independent of the past return $r(t-\tau)$. We expect $\sigma_0(t)$ is a slow-varying and well-behaved function of $t$ since the stochasticity or irregularity of $\sigma(t)$ has been ``averaged" out. One notices that $\sigma_0(t)$, being an expectation value, is not directly observed, since one can not distinguish which part of the observed fluctuation of the return $r(t)$ is due to $\sigma(t)$ and which part is due to $\epsilon(t)$, so we can not have observations of only $\sigma(t)$ at each time, we only know $r(t)$ which is the product of two random variables $\sigma(t)$ and $\epsilon(t)$. In a summary, we can write a general expression of a stock return as follows.
\begin{equation}
r(t) = \sigma(t)\cdot \epsilon(t); \quad \text{with} \quad \e \epsilon(t) = 0,\, \e \epsilon(t) \epsilon(t') = \delta_{t t'} \, \text{and} \,\e \sigma(t) = \sigma_0(t) 
\label{genr}
\end{equation}
where $\delta_{tt'}$ is Kronecker's delta function. \eq{genr}, valid in both ARCH-like and stochastic volatility models, is our starting point. If the volatility is deterministic and constant, $\sigma(t) = \sigma_0$, and $\epsilon(t)$ follows standard normal distribution, we get back the Brownian motion model. To take into account the clustered volatility, the randomness is introduced to the volatility $\sigma(t)$. We follow the approach of ARCH-GARCH models and assume that the randomness is all from the past returns. Thus, we can write $\sigma(t)$ as a deterministic function of past returns.
\begin{equation}
\sigma(t) = \sigma_0(t)+J(t; \{r(t-\tau),\tau>0\}); \quad \e J(t; \{r(t-\tau),\tau>0\}) = 0
\label{genJ}
\end{equation}

the generalized function $J \equiv J(t; \{r(t-\tau),\tau>0\})$ encapsulates all possible complex dependence on the past returns. Because of this past dependence, the volatility at different time becomes correlated, which results in clustered volatility. Some general properties of the return $r(t)$ immediately follow from \eq{genr} and \eqref{genJ}. For example, with $\tau \geq 1$
\begin{equation}
\begin{split}
\e r(t)r(t-\tau) 
&= \e r(t-\tau) (\e \sigma(t) \epsilon(t) | \mathcal{F}_{t-1})\\
&= \e r(t-\tau) \sigma(t) (\e \epsilon(t) | \mathcal{F}_{t-1}) = 0
\end{split}
\end{equation} 
where $| \mathcal{F}_{t-1}$ means conditioning on information available before time $t$ or information up to time $t-1$. As a consequence, the returns at different time are still uncorrelated. Working in a similar way, we see that the observed volatility $\sqrt{\e r^2(t)}$ is larger than $\sigma_0(t)$ (note $\e J = 0$ by \eq{genJ}),
\begin{equation}
\e r^2(t) = \sigma^2_0(t) + \e J^2
\end{equation}
by introducing the randomness to the volatility, the observed volatility increases.

We next consider the specific form of $J$ with the constraint $\e J = 0$.
We further assume the stationary condition, namely $\sigma_0(t) = \sigma_0$ is constant and $J$ does not explicitly depend on time $t$. If $J$ is a linear function of past returns,
\begin{equation}
J(\{r(t-\tau)\}) = \sum_{\tau = 1}^{\infty} K(\tau) r(t-\tau)
\label{eqk}
\end{equation}

then it actually belongs to one kind of QARCH model. Indeed, we can calculate the conditional expectation $\sigma^2_t = \e (r^2(t) | \mathcal{F}_{t-1})$, 
\begin{equation}
\sigma^2_t = \sigma_0^2 + 2 \sigma_0 \sum_{\tau = 1}^{\infty} K(\tau) r(t-\tau) + \sum_{\tau, \tau' = 1}^{\infty} K(\tau) K(\tau') r(t-\tau) r(t-\tau')
\end{equation}
Comparing with \eq{qr1}, we see the above linear model belongs to QARCH type models with kernels $\mathcal{L}(\tau) = 2 \sigma_0 K(\tau)$ and $\mathcal{K}(\tau,\tau') = K(\tau)K(\tau')$. This example shows the simple form of our function $J$ may lead to a more complicated structure of the conditional variance which plays a central role for many ARCH-GARCH models. 

As indicated by leverage effect, the current volatility depends differently on whether the past returns are positive or negative. Instead of the linear function \eq{eqk}, it is more reasonable to propose the asymmetric volatility feedback as follows.
\begin{equation}
\begin{split}
r(t) &= \sigma(t) \epsilon(t) = [\sigma_0 + J(\{r(t-\tau),\tau>0\})] \cdot \epsilon(t) \\
J(\{r(t-\tau),\tau >0\}) &= \sum_{\tau = 1}^{\infty} K_{+}(\tau) (r_{+}(t-\tau)- \e r_{+} ) + \sum_{\tau = 1}^{\infty} K_{-}(\tau) (r_{-}(t-\tau)- \e r_{-})
\end{split}
\label{eqkpm}
\end{equation}

where $r_{+}(t) = r(t)\cdot I(r(t)>0)$ and $r_{-}(t) = r(t)\cdot I(r(t)<0)$, $I(\e)$ is an indicator function, $I(\e) = 1$ if the event $\e$ is true and $I(\e) = 0$ otherwise. $\e r_{\pm} = \e r_{\pm}(t)$, are constants by the stationary assumption. For simplicity, we assume $\epsilon(t)$ is drawn from the standard normal distribution, although other distributions with fat tails can be considered. If $K_+(\tau) = K_-(\tau)$, then the above nonlinear model reduces to the linear case \eq{eqk}. The $\sigma(t)$ is deterministic when the past history and $K_{\pm}(\tau)$ are given. From the general properties, we have
\begin{equation}
\e r(t) = 0; \quad \e r(t)r(t') = \sigma^2 \delta_{t t'}
\end{equation} 
$\sigma^2 = \e r^2(t)$ is the observed variance. $J$ in \eq{eqkpm} can be viewed as a first order expansion with respect to $r_+(t-\tau)$ and $r_-(t-\tau)$ around $\e r_+$ and $\e r_-$. By using $r(t) = r_{+}(t)+r_{-}(t)$ and $|r(t)| = r_{+}(t)-r_{-}(t)$, we can rewrite $J$ as 
\begin{equation}
\begin{split}
J &= \sum_{\tau = 1}^{\infty} K_V(\tau) (|r(t-\tau)|- \e |r|) + \sum_{\tau = 1}^{\infty} K_L(\tau) r(t-\tau)\\
K_V(\tau) &= (K_{+}(\tau)-K_{-}(\tau))/2 ; \quad  K_L(\tau) = (K_{+}(\tau)+K_{-}(\tau))/2
\end{split}
\label{eqkvl}
\end{equation}

$K_V(\tau)$ characterizes the effect on the current volatility from the magnitude of the return at $\tau$ units of time ago. It is very similar to the ARCH model where the conditional volatility depends on the absolute value of the past returns. $K_L(\tau)$ gives $J$ the dependence of the sign of the past returns, which provides the asymmetric behavior. We notice that if $K_+(\tau) =-K_-(\tau)$, then $K_L(\tau)=0$. $J$ only depends on the absolute value of past returns. Similar terms like $K_L(\tau)$ are also included in some GARCH-like models to take into account the leverage effect. However, if one treats $r_{+}(t-\tau)$ and $r_{-}(t-\tau)$ as independent variables, both clustered volatility and leverage effect arise naturally. 

Let us further comment on the relationship between our model and existing volatility models. Our model starting from $\sigma_t$ instead of the variance $\sigma^2_t$ is very similar to the volatility models studied by Taylor\cite{Taylor}, Schwert\cite{Schwert1,Schwert2} and Zakoian\cite{Zakoian}. Among them, TGARCH model studied by Zakoian\cite{Zakoian} is the closest one. In his paper, he performed a careful statistical test on a specific model called TARCH(5) and argued that the different effect between positive and negative past returns is significant. The model written in our notation is
\begin{equation}
\sigma_t = \alpha_0 + \sum_{i=1}^{5} \alpha_{\tau}^{+} r_{+}(t-\tau) -\sum_{i=1}^{5} \alpha_{\tau}^{-} r_{-}(t-\tau)
\end{equation}
which is equivalent (up to a constant) to our model truncated at $\tau_{max} = 5$. Instead of thorough statistical tests on the model, we proceed in a different way to study the model and calibrate it with the data. We do not constrain the model to some small lags, but instead allow it to be large and see what patterns of model parameters like $\alpha_{\tau}^{\pm}$ appear when the model is compared to the data. We hope our method will reveal new aspects about the asymmetric ARCH model and provide useful insights of the volatility process.

\section{Model Calibration in the Leading Order}
In order to calibrate the model and obtain the information about parameters $K_{\pm}(\tau)$, we define various correlation functions as ``observables" that can be derived from the data, then compute them within our model in terms of $K_{\pm}(\tau)$. We study the following observables.
\begin{equation}
\begin{split}
L_{+}(t,\tau) &= \e r^2(t) r_{+}(t-\tau) - \e r^2(t)\cdot \e r_{+}(t-\tau)\\
L_{-}(t,\tau) &= \e r^2(t) r_{-}(t-\tau) - \e r^2(t)\cdot \e r_{-}(t-\tau)\\
L(t,\tau) &= \e r^2(t) r(t-\tau) = L_{+}(t,\tau)+L_{-}(t,\tau)\\
V(t,\tau) &= \e r^2(t) r^2(t-\tau) - \e r^2(t)\cdot \e r^2(t-\tau)
\end{split}
\label{obs}
\end{equation}

Under the stationary assumption, the above functions are independent of time $t$ and only depend on $\tau$. The observable $L(\tau) = L(t,\tau)$, up to a normalization factor, was first proposed by Bouchaud, Matacz and Potters\cite{reBouchaud} to study the leverage effect. The leverage effect is characterized by the following properties of $L(\tau)$.
\begin{equation}
L(\tau) = 
\begin{cases}
\, 0 \, , & \tau <0 \\
\, <0 \, ,& \tau \geq 0 \\
\end{cases}
\end{equation}
This asymmetry in time is a built-in feature in our framework, since the current volatility depends on past returns not future.

To further explore the consequences of our model \eq{eqkpm}, we need to evaluate the correlation functions defined in \eq{obs}. The expression $J$ in \eq{eqkpm} can be regarded as the first order Taylor expansion of $r_{\pm}$ and expect $K_{\pm}(\tau)$ to be small in some sense, for example, $K_{\pm}(\tau) < 1$. In calculating the correlation functions, we only keep the first order of $K_{\pm}(\tau)$. Due to the smallness of both $K_{\pm}(\tau)$ and $r(t)$, higher order corrections are expected to be suppressed. We will discuss later whether the first order calculation is reasonable by retrospecting the results obtained. In the leading order of $K_{\pm}(\tau)$, $L_{+}(\tau)$ can be expressed as
\begin{equation}
\begin{split}
L_{+}(\tau) &= \e (\sigma_0^2 + 2\sigma_0 J + J^2)r_{+}(t-\tau) - (\sigma_0^2+\e J^2)\cdot \e r_{+}(t-\tau) \\
& \approx 2 \sigma_0 \e J r_{+}(t-\tau) \\
& =2 \sigma_0  \sum_{\tau' = 1}^{\infty} K_{+}(\tau') (\e r_{+}(t-\tau')r_{+}(t-\tau) - \e r_{+} \e r_{+}) \\
& + 2 \sigma_0 \sum_{\tau' = 1}^{\infty} K_{-}(\tau') (\e r_{-}(t-\tau')r_{+}(t-\tau) - \e r_{-} \e r_{+}) 
\end{split}
\end{equation}
Note the expectation values in parenthesis are multiplied by $K_{\pm}(\tau)$. To calculate $L_{+}(\tau)$ in leading order, we can use $r(t) = \sigma_0 \epsilon(t)$ in evaluating those expectation values, any deviations from this result are in higher orders of $K_{\pm}(\tau)$. Thus,
\begin{equation}
\begin{split}
\e r_{+}(t-\tau')r_{+}(t-\tau) - \e r_{+} \e r_{+} &\approx \sigma_0^2 \cdot (\e \epsilon_{+}(t-\tau') \epsilon_{+}(t-\tau) - \e \epsilon_{+} \e \epsilon_{+})\\
& = \sigma_0^2 (\e \epsilon^2_{+}-(\e \epsilon_{+})^2) \delta_{\tau \tau'}\\
& = \frac{1}{2}\left(1-\frac{1}{\pi} \right) \sigma_0^2 \delta_{\tau \tau'}
\end{split}
\end{equation}
where $\epsilon_{+}(t) = \epsilon(t)I(\epsilon(t)>0)$ and $\epsilon_{-}(t) = \epsilon(t)I(\epsilon(t)<0)$, we have used $\epsilon(t)$ is drawn independently from the standard normal distribution. Similarly, 
\begin{equation}
\begin{split}
\e r_{-}(t-\tau')r_{+}(t-\tau) - \e r_{-} \e r_{+} &\approx \sigma_0^2 \cdot (\e \epsilon_{-}(t-\tau') \epsilon_{+}(t-\tau) - \e \epsilon_{-} \e \epsilon_{+})\\
& = \sigma_0^2 (\e \epsilon_{+})^2 \delta_{\tau \tau'} \\
& = \frac{1}{2 \pi} \sigma_0^2 \delta_{\tau \tau'}
\end{split}
\end{equation}
where $\epsilon_{+}(t)\cdot \epsilon_{-}(t) =0$ and $\e \epsilon_{-} = - \e \epsilon_{+}$ are used. We have the expression for $L_{+}(\tau)$.
\begin{equation}
L_{+}(\tau) = \sigma_0^3\left[ (1-\frac{1}{\pi}) K_{+}(\tau) + \frac{1}{\pi}K_{-}(\tau) \right]
\label{L+}
\end{equation}

Similarly for $L_{-}(\tau)$,
\begin{equation}
L_{-}(\tau) = \sigma_0^3\left[ \frac{1}{\pi} K_{+}(\tau) + (1-\frac{1}{\pi})K_{-}(\tau) \right]
\label{L-}
\end{equation}

The overall scale of $L_{\pm}(\tau)$ is given by $\sigma_0^3$. We can approximate $\sigma^2_0$ with the observed variance $\sigma^2$ in the first order since $\sigma^2 = \sigma_0^2 + \e J^2$ and any corrections come from the second order. The different weighting values ($1/\pi$ and $1-1/\pi$) in \eq{L+} and \eqref{L-} introduce the asymmetry between $L_+(\tau)$ and $L_-(\tau)$. Having obtained $L_{\pm}(\tau)$, we have the expression for the leverage function.
\begin{equation}
L(\tau) = L_+(\tau)+L_-(\tau) = 2\sigma_0^3 K_L(\tau) \approx 2\sigma^3 K_L(\tau)
\label{L}
\end{equation}

The leverage function $L(\tau)$ (up to a factor $\sigma^4$) was first studied in the paper\cite{reBouchaud}. The authors argue that if the current price is determined by some average of the past prices and assume the volatility is proportional to the current price, then one obtains a universal relationship $\lim_{\tau \to 0} L(\tau)/\sigma^4 = -2 $. In our model, this is true only when $K_L(\tau \to 0) \sim -\sigma$. In the viewpoint of our model, the universal constant $-2$ means the scale of the parameter $K_L(\tau)$ with small $\tau$ is set by $-\sigma$. This scale is quite understandable since one expects the leverage effect is more significant when there is a large downward move in the market which induces a large volatility. Thus, $K_L(\tau)$ is likely to scale with $\sigma$. We do not further assume any postulated properties about $K_L(\tau)$ or $K_{+}(\tau)$ and $K_{-}(\tau)$, instead $K_{\pm}(\tau)$ are derived from the observables $L_{\pm}(\tau)$ by \eq{L+} and \eqref{L-}.
\begin{equation}
\begin{split}
K_{+}(\tau) &= \frac{(\pi-1)L_{+}(\tau)- L_{-}(\tau)}{\sigma^3(\pi-2)}\\
K_{-}(\tau) &= -\frac{L_{+}(\tau)-(\pi-1) L_{-}(\tau)}{\sigma^3(\pi-2)}\\
\end{split}
\label{Kpm}
\end{equation}

The above equations completely fix our model parameters $K_{+}(\tau)$ and $K_{-}(\tau)$, then they can be used to calculate other observables and compare with the data. In this way, we can have a consistency check of our model as well as leading order calculation. Let's calculate $V(\tau) = V(t, \tau)$ that is usually used to illustrate the volatility clustering. 
\begin{equation}
\begin{split}
V(\tau) &= \e (\sigma_0^2 + 2 \sigma_0 J + J^2) r^2(t-\tau) - (\sigma_0^2 + \e J^2)\e r^2(t-\tau)\\
&\approx 2 \sigma_0 \e J r^2(t-\tau)\\&=2 \sigma_0  \sum_{\tau' = 1}^{\infty} K_{+}(\tau') (\e r_{+}(t-\tau')r^2(t-\tau) - \e r_{+} \e r^2) \\
& +2 \sigma_0 \sum_{\tau' = 1}^{\infty} K_{-}(\tau') (\e r_{-}(t-\tau')r^2(t-\tau) - \e r_{-} \e r^2)\\
& \approx 2 \sigma_0^4 \left[ K_{+}(\tau) (\e \epsilon^3_{+}-\e \epsilon_{+}) +K_{-}(\tau)(\e \epsilon^3_{-}-\e \epsilon_{-}) \right]\\
& = \sqrt{\frac{2}{\pi}}\sigma_0^4(K_{+}(\tau)-K_{-}(\tau))
\end{split}
\end{equation}
Thus,
\begin{equation}
V(\tau) = 2 \sqrt{\frac{2}{\pi}}\sigma^4 K_V(\tau)
\label{V}
\end{equation}

We have seen from \eq{L} and \eq{V} that $K_L(\tau)$ and $K_V(\tau)$ are indeed responsible for the leverage effect and clustered volatility respectively.

We can calculate the even moments of the return $r(t)$, which provides another aspect to check our model and the first order calculation. Define the normalized even moment function $M_n$ as follows.
\begin{equation}
M_n = \frac{\e r^n(t)}{(\e r^2)^{n/2}}, \quad \text{n is even}
\label{Mndef}
\end{equation}

In the Gaussian case, namely $r(t) = \sigma_0 \epsilon(t)$, the expression of $M_n$ is easily obtained.
\begin{equation}
M_{n0} = \e \epsilon^n = \frac{2^{n/2}}{\sqrt{\pi}} \Gamma(\frac{n+1}{2}) = 1\cdot 3 \cdot 5 \cdot \cdots (n-3) \cdot (n-1)
\label{Mn0}
\end{equation}
 
To calculate $M_n$ in the leading nonzero order of $K_{\pm}(\tau)$, we first evaluate $\e r^2(t)$.
\begin{equation}
\begin{split}
\e r^2(t) &= \sigma_0^2 + \e J^2\\
& = \sigma_0^2 + \sum_{\tau,\tau' = 1}^{\infty}K_{+}(\tau) K_{+}(\tau') (\e r_{+}(t-\tau) r_{+}(t-\tau') -\e r_+ \e r_+)\\
& + \sum_{\tau,\tau' = 1}^{\infty}K_{-}(\tau) K_{-}(\tau') (\e r_{-}(t-\tau) r_{-}(t-\tau') -\e r_- \e r_-)\\
& + 2 \sum_{\tau,\tau' = 1}^{\infty}K_{+}(\tau) K_{-}(\tau') (\e r_{+}(t-\tau) r_{-}(t-\tau') -\e r_+ \e r_-)\\
& \approx \sigma_0^2 \cdot \biggl[ 1+ (\e \epsilon^2_{+} - (\e \epsilon_{+})^2) \left( \sum_{\tau = 1}^{\infty} K^2_+(\tau) + K^2_-(\tau)\right)+ 2 (\e \epsilon_{+})^2 \sum_{\tau = 1}^{\infty} K_+(\tau) K_-(\tau) \biggr]\\
& = \sigma_0^2\cdot \left[ 1+ \frac{1}{2}(1-\frac{1}{\pi}) \sum_{\tau = 1}^{\infty}\left(K_+(\tau) + K_-(\tau)\right)^2+(\frac{2}{\pi}-1) \sum_{\tau = 1}^{\infty} K_+(\tau) K_-(\tau) \right]\\
& = \sigma_0^2 (1+\Delta(K))
\end{split}
\end{equation}
where 
\begin{equation}
\Delta(K) = \frac{1}{2}(1-\frac{1}{\pi}) \sum_{\tau = 1}^{\infty}\left(K_+(\tau) + K_-(\tau)\right)^2+(\frac{2}{\pi}-1) \sum_{\tau = 1}^{\infty} K_+(\tau) K_-(\tau)
\end{equation}
is the leading nonzero correction to the $\e r^2(t)$, or $\e J^2 \approx \sigma_0^2 \Delta(K)$. Now, let's calculate $\e r^n(t)$ for even $n \geq 2$ in the leading nonzero order of $K$, or equivalently, in the leading order of $\Delta(K)$.
\begin{equation}
\begin{split}
\e r^n(t) &= \e \epsilon^n \cdot (\sigma_0^n + \frac{n(n-1)}{2}\sigma_0^{n-2} \e J^2 +...)\\
& \approx \sigma^n_0 \e \epsilon^n \cdot \left(1+\frac{n(n-1)}{2} \Delta(K) \right)
\end{split}
\end{equation}
Therefore, keep the first order of $\Delta(K)$ in $M_n$.
\begin{equation}
\begin{split}
M_{n} &\approx M_{n0}(1-\frac{n}{2}\Delta(K)) \left(1+\frac{n(n-1)}{2} \Delta(K) \right)\\
&\approx M_{n0} \left( 1+\frac{n(n-2)}{2} \Delta(K) \right)
\end{split}
\label{Mn1}
\end{equation}

We have worked out the observables $L_{\pm}(\tau)$, $V(\tau)$ and $M_n$ in terms of functions $K_{\pm}(\tau)$ at the leading order. The simple leading order results allow us to easily calibrate our model and interpret the data.

\section{Empirical Study: a stationary case}
To perform an empirical study of the model, we analyze daily closing prices of the 30 stocks forming the Dow Jones Industrial Average over 77 years from 15 March 1939 until 16 October 2015. There are jumps in those stocks' prices such as stock splits, bonus issues, dividends payouts and replacements of index components. We simply discard those points by setting the return zero if the price jumps by more than $15\%$ in one day. This regularization condition should not introduce much bias in our analysis, since they take up only $0.1\%$ of our total 19800 data points for one stock. We also change this cutoff in a range $10\% \sim 25\%$, our conclusion remains the same. For this section, we work under the stationary assumption and assume function $K_{\pm} (\tau)$ only depends on the time lag $\tau$. We will relax the stationary assumption and study how $K_{\pm}$ change with time in the next section. The time average of data in the whole period is used to estimate the different expectation values defined in the previous section. The stationary assumption across 77 years is certainly over-simplified, however we expect the qualitative behavior of functions $K_{\pm}$ do not change much. In this section, we will focus on how $K(\tau)$ depends on $\tau$ and check our simple model against Dow Jones data. 

\begin{figure}[h!]
\includegraphics[width = 130mm, height = 80mm]{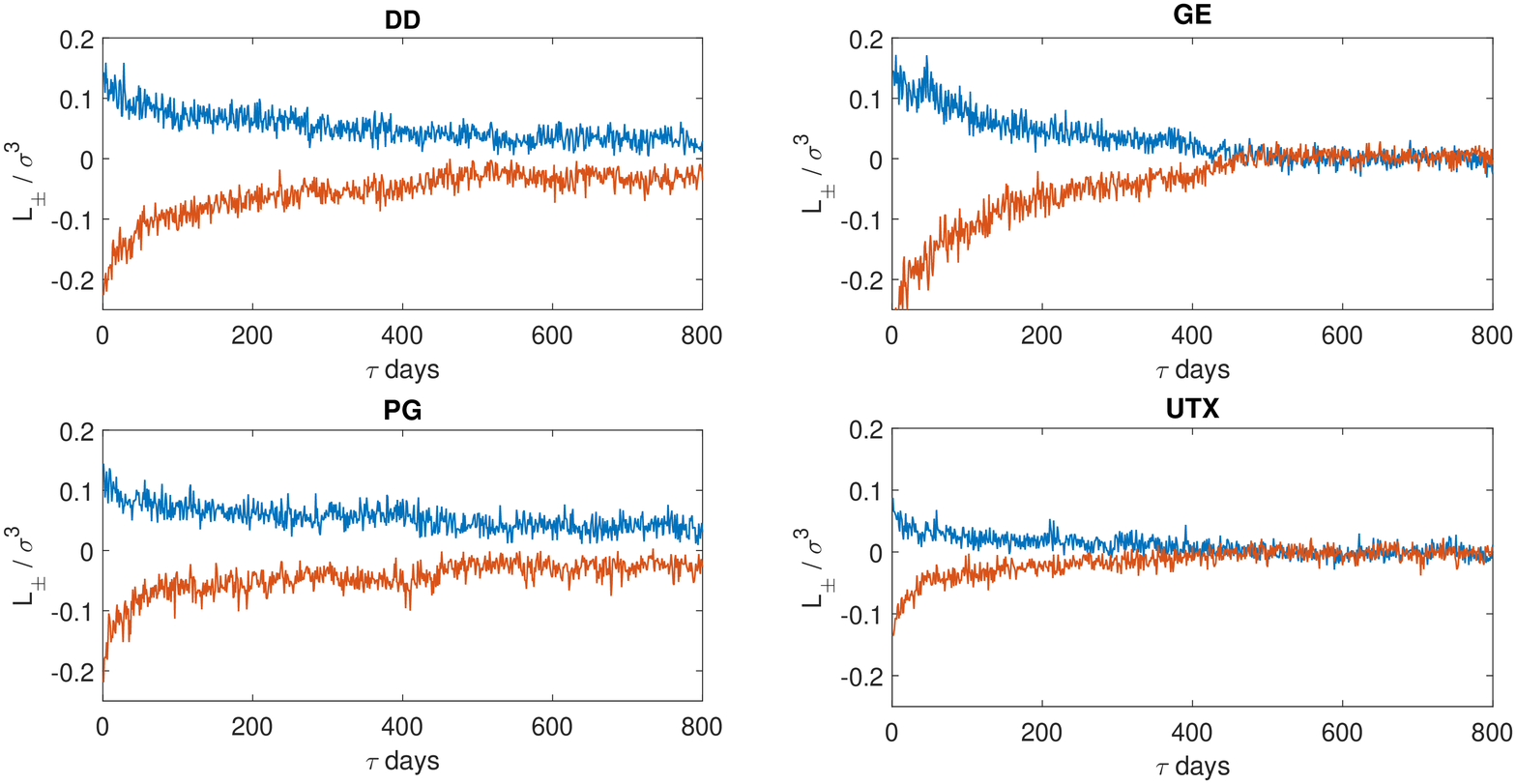}
\caption{$L_{\pm} (\tau) / \sigma^3$ for different stocks.}
\end{figure}

Fig.1 shows $L_{\pm} (\tau)$ for different stocks. We choose the 4 stocks with ticker symbols DD, GE, PG and UTX since they are only stocks remaining in Dow Jones Industrial Average for the whole period under studied. We notice 3 major features of the conditional correlation functions $L_{\pm}(\tau)$. First, $L_-(\tau) \approx -L_+(\tau)$, especially for large $\tau$. This fact shows $K_-(\tau) \approx -K_+(\tau)$, which means $K_V(\tau) \gg K_L(\tau)$. The volatility feedback largely depends on the magnitude of the past returns not linearly on the returns itself. The leverage effect is small as compared to the correlation on volatility. Second, $L_-(\tau)$ and $L_+(\tau)$ indeed behave differently at small $\tau$. For example, $|L_{-}(\tau = 1)|>L_{+}(\tau = 1)$ and $L_-(\tau)$ decays slightly faster than $L_+(\tau)$ does. This remarkable difference at short time scale is the evidence that the leverage effect although being small is significant and operating strongly at short time. Third, $L_{\pm} (\tau)$ are different for different stocks. For some stocks like GE and UTX, the correlation functions decay faster than the correlation functions of other stocks like PG and DD. This difference may be due to the fact that stocks are from different sectors. 

Given $L_{\pm}(\tau)$, we can obtain $K_{\pm}(\tau)$ by \eq{Kpm}. We choose UTX as an example to study the functions $K_{\pm}(\tau)$ in detail since $L_{\pm}(\tau)$ of UTX is relatively small so that the result from the first order calculation is more reliable.

\begin{figure}[h!]
\includegraphics[width = 130mm, height = 80mm]{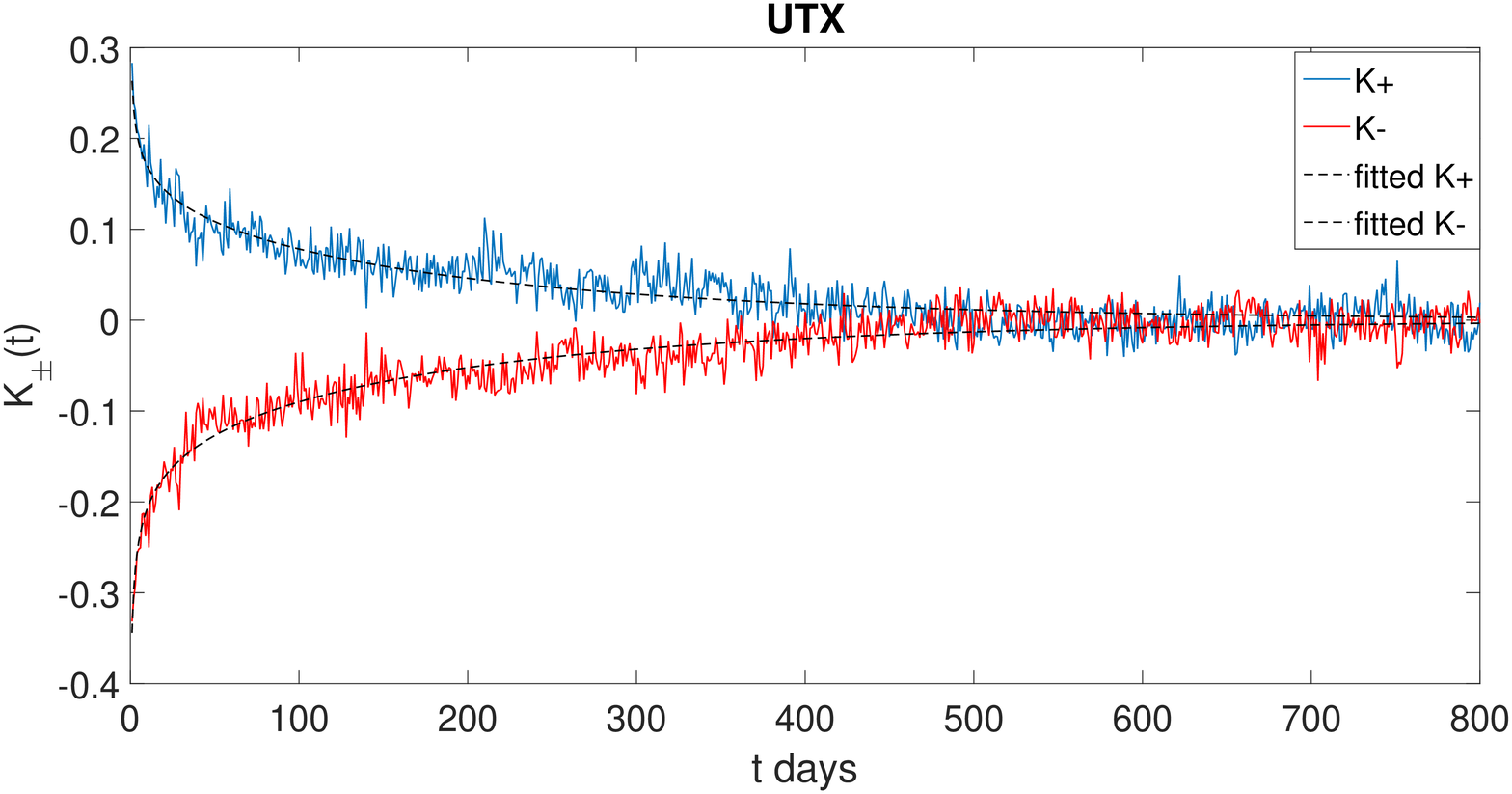}
\caption{$K_{\pm} (\tau)$ for UTX.}
\end{figure}

Fig.2 illustrates the behaviors of function $K_{\pm}(\tau)$. To account both fast and slow decays of correlation functions, we propose the exponential truncated power-law fitting functions for $K_{\pm}(\tau)$ as follows.
\begin{equation}
K(\tau) = \frac{A}{\tau^{b}} e^{-\tau/T}
\end{equation}
when $b = 0$, it is an exponential decay which is more suitable for fitting a stock like GE; when $T \to \infty$ or $T > \tau_{max}$ ($\tau_{max} = 800$ in our case), it is a power-law decay which works for a stock like PG. 

The fitted functions for UTX are (with $95\%$ confidence bounds)
\begin{equation}
\begin{split}
K_{+}(\tau) &= (0.26 \pm 0.02) \cdot \tau^{-0.17 \pm 0.02} e^{-\tau/(245 \pm 20)} \\ 
K_{-}(\tau) &= (-0.34 \pm 0.02) \cdot \tau^{-0.20 \pm 0.02} e^{-\tau/(247 \pm 20)}
\end{split}
\label{fitK}
\end{equation}

The fitting parameters $A_{\pm}$, $b_{\pm}$ and $T_{\pm}$ illustrate the different behaviors of $K_{\pm}(\tau)$ quantitatively. We have seen that $|A_-|>A_+$ and $b_- > b_+$ while $T_- \approx T_+$. $|A_-|>A_+$ shows the feedback to the current volatility is stronger when the past returns are negative. The contribution of $b$ to the function $K(\tau)$ is through the factor $\tau^{-b}$. For small $\tau$, the value of $\tau^{-b}$ is very sensitive to the parameter $b$. The parameter $b$ characterizes the feedback effect at short time scale. The larger the $b$ is, the faster $|K(\tau)|$ drops. The parameter $T$ gives us the long-term time scale of the correlation. This long term correlation is well known and results in the clustered volatility, which has a similar time scale for both feedback functions $K_{\pm}(\tau)$. Therefore, we have two time scales on the feedback effect of volatility. To estimate the two time scales, we can alternatively fit $K(\tau)$ with two exponential functions.
\begin{equation}
\begin{split}
K_{+}(\tau) &= 0.14 \, e^{-\tau/9} + 0.13 \, e^{-\tau/200}\\ 
K_{-}(\tau) &= -0.18 \, e^{-\tau/12} - 0.15 \, e^{-\tau/200}
\end{split}
\end{equation}
We see that for UTX the short time scale is about 10 days while the long time scale is 200 days. Doing similar two exponential fitting to other stocks yields the short time scale is typically $10 \sim 50$ days while the long time scale can be $300 \sim 1000$ days. The two time scales are consistent with the two time scales found in Chicheportiche and Bouchaud's study of the QARCH models\cite{qrBouchaud1}. 

To further understand the features of quadratic kernel in QARCH model, we square \eq{eqkvl}.
\begin{equation}
\sigma_t^2 = \cdots \sum_{\tau = 1}^{\infty} (K^2_V(\tau)+K^2_L(\tau)) r^2(t-\tau) + \sum_{\tau \neq \tau'}^{\infty}K_L(\tau)K_L(\tau')r(t-\tau) r(t-\tau') + \cdots
\label{qr2}
\end{equation}

The terms omitted are constant terms, linear terms in $r(t-\tau)$ or $|r(t-\tau)|$, and quadratic terms like $r(t-\tau) |r(t-\tau')|$ and $|r(t-\tau)| |r(t-\tau')| (\tau \neq \tau')$. If one instead assumes a QARCH model \eq{qr1}, the correlation functions or generalized moments such as $\e r^2(t) r(t-\tau)$ or $\e r^2(t) r(t-\tau) r(t-\tau')$ will be used to calibrate the QARCH model. This is very similar to the way the QARCH model was calibrated in the paper\cite{qrBouchaud1}, where the combined generalized moment methods(GMM) and maximum likelihood estimation(MLE) were used. However, if one assumes our model \eq{eqkvl}, in the leading order of $K_{\pm}(\tau)$, only two terms appear in \eq{qr2} will contribute to $\e r^2(t) r(t-\tau) r(t-\tau')$. Therefore, we have an estimate of quadratic kernel in our model as
\begin{equation}
\mathcal{K}(\tau,\tau') \approx
\begin{cases}
\, K^2_V(\tau)+K^2_L(\tau), & \quad \tau = \tau' \\
\, K_L(\tau)K_L(\tau'), & \quad \tau \neq \tau'
\end{cases}
\end{equation}       
with the help of above equation, the features of the $\mathcal{K}(\tau,\tau')$ can be understood in terms of $K_V(\tau)$ and $K_L(\tau)$. Because $K_V(\tau) \gg |K_L(\tau)|$, the diagonal terms in the quadratic kernel of QARCH model are larger than its off-diagonal terms. This because the $\sigma_t$ largely depends on $|r(t-\tau)|$ not $r(t-\tau)$ itself, and the contribution of $|r(t-\tau)|$ can enter into the diagonal elements of $\mathcal{K}(\tau,\tau')$ but not its off-diagonal terms. It also can be understood that off-diagonal terms may become negative for large lag $\tau$. Indeed, $K_{+}(\tau) \sim -K_{-}(\tau)$ for a large $\tau$ and $K_L(\tau) = (K_{+}(\tau)+K_{-}(\tau))/2$, so the signs of $K_L(\tau)$ and $K_L(\tau')$ can well be opposite, which renders some off-diagonal elements negative.  

We have obtained the functions $K_{\pm}(\tau)$ from the conditional correlation functions $L_{\pm}(\tau)$, then we can use the fitted functions  \eq{fitK} to compute other observables.

\begin{figure}[h!]
\includegraphics[width = 120mm, height = 60mm]{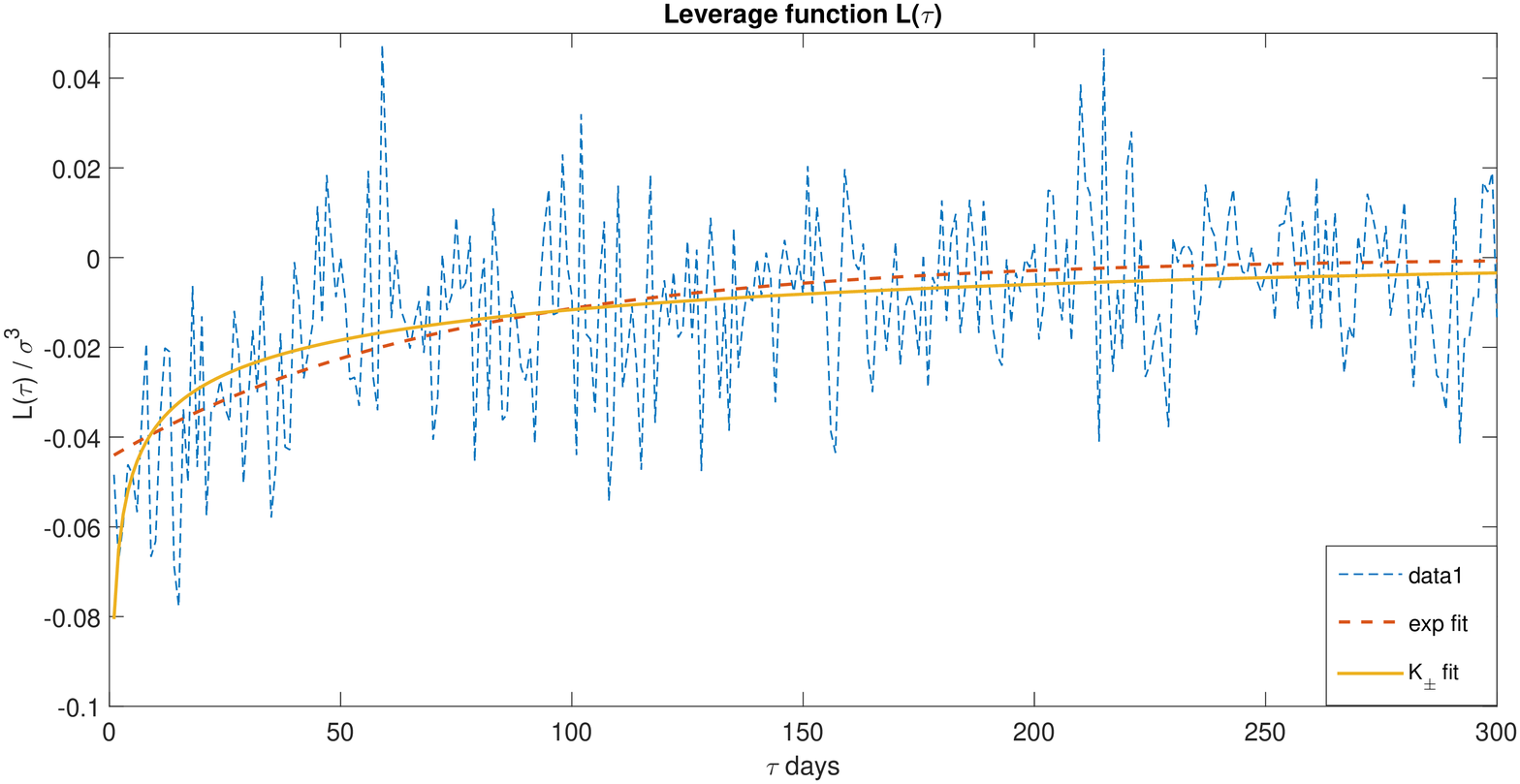}
\caption{ Compare the fits for the leverage function $L(\tau)$.}
\end{figure}

Fig.3 compares two fits with the leverage function $L(\tau)$. It is a good fit for the data without surprising since $L(\tau) = L_-(\tau)+L_+(\tau)$ and $K_{\pm}(\tau)$ are fitted from $L_{\pm}(\tau)$ respectively, however it indeed shows that the direct one exponential fit for $L(\tau)$, as was usually done in previous literature\cite{reBouchaud,SV}, underestimates the leverage effect for individual stocks at short time. In fact, the leverage effect at $\tau = 1$ is twice as large as the result from the one exponential fit. The large noise in $L(\tau)$ also suggests that it is better to use $L_{\pm}(\tau)$ to study leverage effect and calibrate the model. 

\begin{figure}[h!]
\includegraphics[width = 110mm, height = 60mm]{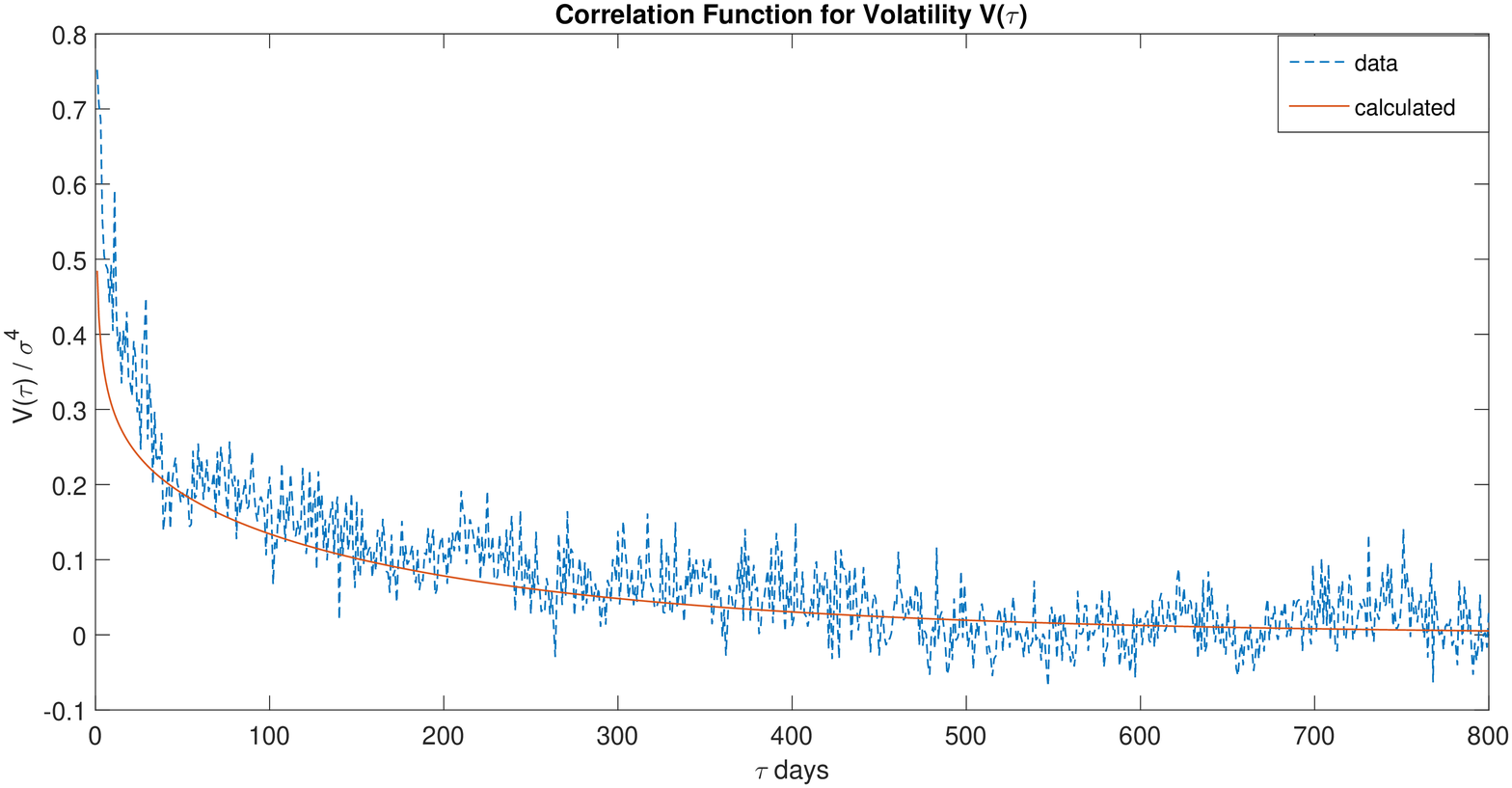}
\caption{ Compare calculated volatility correlation $V(\tau)$ with data }
\end{figure}

Fig.4 shows that $V(\tau)$ is mainly reproduced by the fitted functions. The calculated result only underestimates $V(\tau)$ at small $\tau$ where $K_{\pm}(\tau)$ becomes large and the leading order calculation is less accurate.

\begin{figure}[h!]
\includegraphics[width = 110mm, height = 60mm]{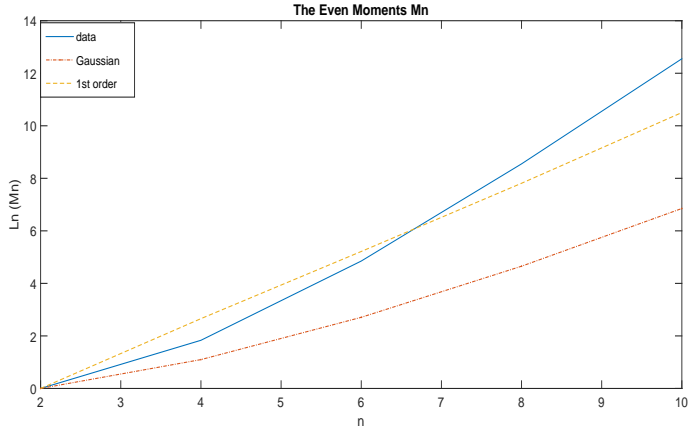}
\caption{ Compare 1st order calculated even moments with the data }
\end{figure}

Fig.5 shows the even moment function $M_n$ up to $n = 10$ and the plot is in logarithmic scale. $M_n$ is only evaluated at $n = 2,4,6,8,10$, the line connected between those values shows how $M_n$ grows with n. We see that the first order result gives much better overall fit to the data than the Gaussian case.

To conclude the empirical study under the stationary assumption, we have calibrated the model in the leading order to the UTX stock returns and checked the first order result is consistent. We find $K_{\pm}(\tau)$ have two time scales. In a short time scale, $\tau = 10 \sim 50$ days, $|K_{-}(\tau)| > K_{+}(\tau)$; In a long time scale, $\tau = 300 \sim 1000$ days, $K_{+}(\tau) \approx -K_{-}(\tau)$. We have checked the consistency of the model and calculation by comparing data with calculated $V(\tau)$ and $M_n$, the result agrees very well with the expectation from perturbation theory.

\section{Empirical Study: a non-stationary case}

It is unlikely that the feedback coefficients $K_{\pm}(\tau)$ are the same during the whole period from 1939 to 2015. In general, they may change from time to time. To assess this time dependent, we calculate the observables like $L_{\pm}(t,\tau)$ at time $t$ by using the data in previous $\Delta t$ days, namely, the data at a small window $[t-\Delta t+1,t]$. We need to pick an appropriate value of $\Delta t$. It should be large enough to draw any statistically meaningful results and small enough to reflect possible time dependent behaviors of the feedback coefficients $K_{\pm}(t, \tau)$. We set $\Delta t = 400$ days, which is comparable to the time scale of clustered volatility. Since now we have a much smaller set of data, we can not reliably estimate $K_{\pm}(t,\tau)$ for large $\tau$ lags, instead we use the average value of first $\Delta \tau$ lags. The averaged $\bar{L}_{\pm}(t)$ and $\bar{K}_{\pm}(t)$ are defined as follows. 
\begin{equation}
\bar{L}_{\pm}(t) = \frac{1}{\Delta \tau}\sum_{\tau = 1}^{\Delta \tau} L_{\pm}(t,\tau), \quad \bar{K}_{\pm}(t) = \frac{1}{\Delta \tau}\sum_{\tau = 1}^{\Delta \tau} K_{\pm}(t,\tau),
\end{equation}
We also need to pick up a reasonable value of $\Delta \tau$. It can not be too large since for a large $\tau$, $K_{\pm}(t,\tau)$ is small but the noise is large. We set $\Delta \tau = 10$ days which is comparable to the time scale of leverage effect. To separate the effects of leverage and clustered volatility, we further define the averaged $\bar{K}_L(t)$ and $\bar{K}_V(t)$ as
\begin{equation}
\bar{K}_L(t) = (\bar{K}_+(t)+\bar{K}_-(t))/2, \quad  \bar{K}_V(t) = (\bar{K}_+(t)-\bar{K}_-(t))/2
\end{equation}
$\bar{K}_L(t)$ and $\bar{K}_V(t)$ measure how leverage effect and clustered volatility changes with time. Those functions are defined for individual stocks. They are in general different from stock to stock. For the non-stationary study, we focus on using those functions as a whole market indicators. Therefore, we need to average $\bar{K}_L(t)$ and $\bar{K}_V(t)$ among different stocks to obtain a representative estimate of the market. The simple arithmetic average is adopted here. Unlike in the stationary case, we do not confine ourselves to study only stocks in the index that survived in the whole period. Our purpose is to estimate the status of the market as a whole and the new stock that replaced the old one comes from the same market that we try to estimate. After all, we average all the 30 stocks in the end. The stock replacements in the index should not affect very much our averaged results. For a comparison, we also analyze a more conventional market indicator, the volatility of the index return. The index return $R_{\text{ind}}(t)$ is defined as follows.

\begin{equation}
R_{\text{ind}}(t) = \text{ln}(I(t+1)/D(t+1)) - \text{ln}(I(t)/D(t)) 
\end{equation}
where $I(t)$ is Dow Jones index value at time $t$ and $D(t)$ is Dow Jones index divisor which makes the index value continuous. The volatility of index return at time $t$, $\sigma_R(t)$, is defined to be the standard deviation of index returns during the time window $[t-\Delta t+1,t]$.  

\begin{figure}[h!]
\includegraphics[width = 140mm, height = 80mm]{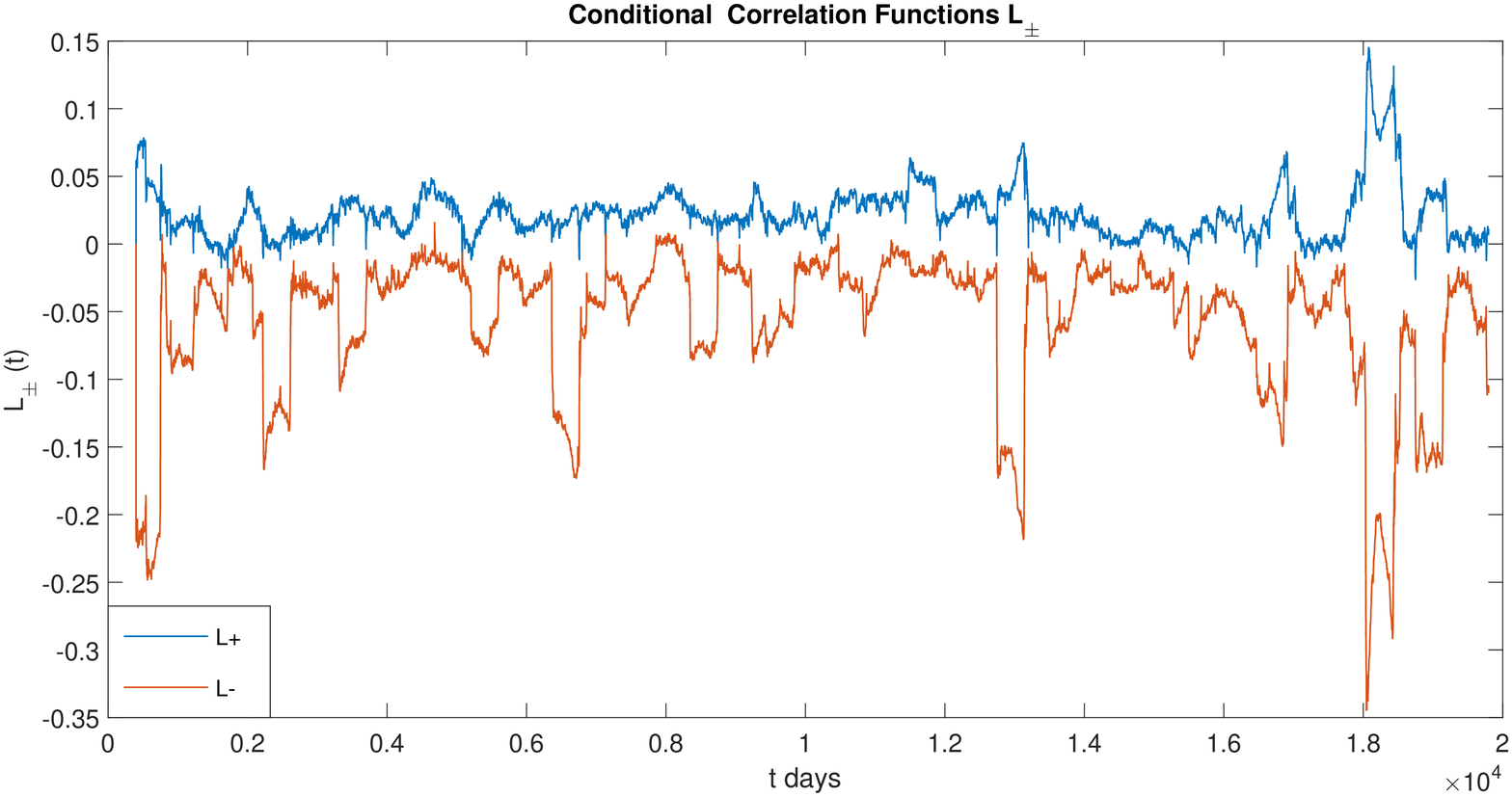}
\caption{ $\bar{L}_{\pm}(t)$ changes with time }
\end{figure}

Fig.6 shows $\bar{L}_-(t)$ and $\bar{L}_+(t)$ are non-stationary and behave very differently. The fluctuation of $\bar{L}_-(t)$ is significantly large than $\bar{L}_+(t)$. This justifies our treatment of regarding $r_{-}(t)$ and $r_{+}(t)$ as independent variables. From $\bar{L}_{\pm}(t)$, we can obtain $\bar{K}_{\pm}(t)$ or $\bar{K}_V(t)$ and $\bar{K}_L(t)$. 

\begin{figure}[h!]
\includegraphics[width = 140mm, height = 80mm]{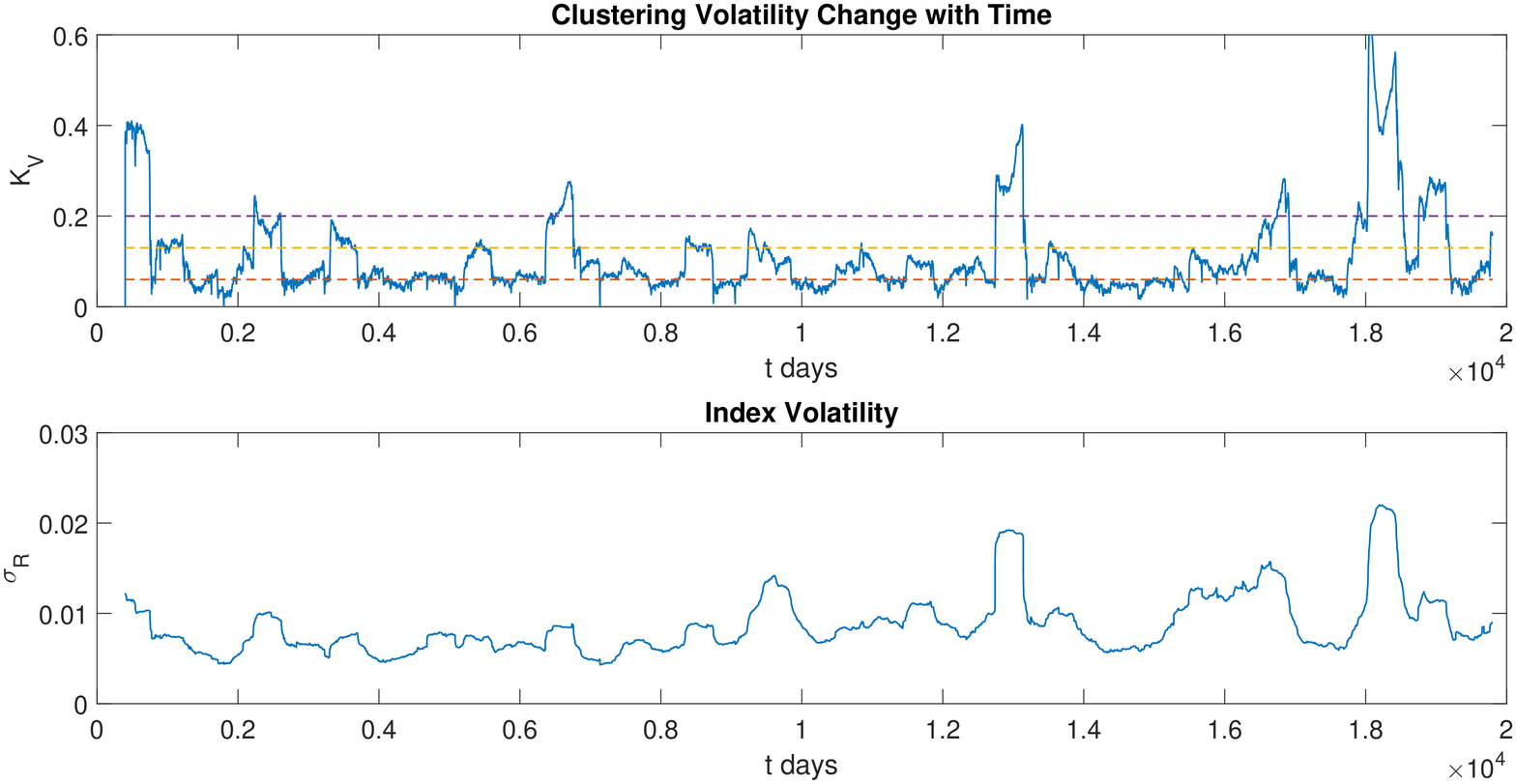}
\caption{ $\bar{K}_V(t)$ and $\bar{\sigma}_R(t)$ change with time }
\end{figure}

Fig.7 shows how $\bar{K}_V(t)$ change with time. It is certainly not constant and it experiences alternative highs and lows. The volatility of the index return in the same moving average window is also plotted in Fig.7. The clustered volatility manifests itself as alternative high and low plateaus. Those plateaus show that we indeed can regard volatility and volatility feedback coefficient $\bar{K}_V(t)$ as constants for a small time window. We further notice that the jumps of $\bar{K}_V(t)$ are more regular than the jumps of index volatility, namely we can roughly separate the value of $\bar{K}_V(t)$ in the whole history by 3 ``phases". As indicated by 3 dashed horizontal lines in Fig.7, the low volatility feedback is $\bar{K}_V(t) \sim 0.06$,  the medium one is $0.13$, and the high one is larger than $0.20$, which normally happened in financial crisis periods. We note $\bar{K}_V(t)$ smaller than $0.6$ and around $0.1$ for the most of the time, which partly justify to regard $K_{\pm}(t,\tau)$ as a small quantity in our perturbative calculation. We notice that the large volatility is usually  accompanied by the large value of $\bar{K}_V(t)$. This implies that the first order feedback is not large enough to explain the appearance of large volatility, so $\bar{K}_V(t)$ itself needs to become large in order to fit the data. A possible way to deal with this large deviation of returns within the model is to assume that the innovation $\epsilon(t)$ is not Gaussian but some heavy tailed distribution. We can still work in the first order of $K_{\pm}(\tau)$, but now some constants like $1-1/\pi$ will be changed according to the specific distribution chosen. Such a modification may make $K_{\pm}(\tau)$ more stable, but it does not change the picture qualitatively.

\begin{figure}[h!]
\includegraphics[width = 140mm, height = 80mm]{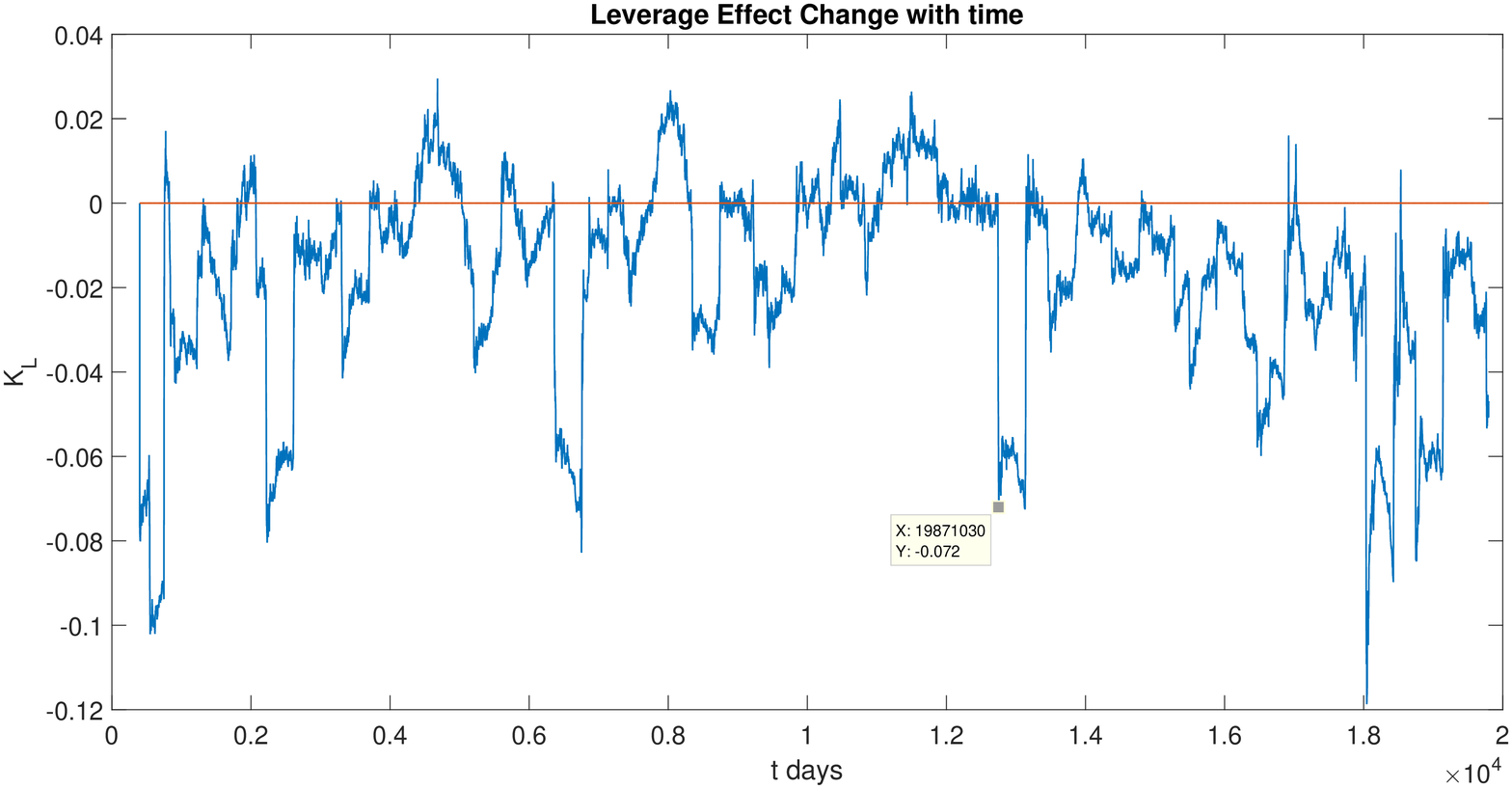}
\caption{ $\bar{K}_L(t)$ change with time }
\end{figure}

Fig.8 depicts the evolution of $\bar{K}_L(t)$. The scale of $\bar{K}_L(t)$ is $-0.08 \sim 0.02$ much smaller than $\bar{K}_V(t)$. Surprisingly, $\bar{K}_L(t)$ is not always negative and it can be positive in some certain of periods. It is only consistently below zero in recent years especially after 1987 crash. Although there are times its value jumps above zero after 1987, it roughly fluctuates around $-0.02$ except for financial crisis. As in the case of $\bar{K}_V(t)$, one can also categorize the market states into 3 ``phases" according to the value of $\bar{K}_L(t)$. This ``phase transition" is most clearly seen in the recent decade. If we plot $\bar{K}_L(t)$ from $2005$ to $2015$ as in Fig.9, one can easily see the pattern switching behavior. Before $2008$ financial crisis, the leverage feedback is small $\bar{K}_L(t) \sim -0.02$. In the period $2008 \sim 2010$ and $2011 \sim 2013$, the leverage feedback is large $\bar{K}_L(t) \sim -0.06$. In $2013 \sim 2014$, $\bar{K}_L(t)$ came back to $-0.02$, but at the end of $2014$, it dropped to a medium value $-0.03$ and stayed here until the summer crash of $2015$, which brought the value down to $-0.06$. The index volatility of the same period is plotted in Fig.9. We have seen the high volatility usually implies the large leverage feedback, but $\bar{K}_L(t)$ contains additional subtle features of the market that are difficult to see in the volatility alone. For example, the volatility in the period $2011 \sim 2013$ is much smaller than that in the period $2008 \sim 2010$, but the leverage feedback is similar in those two periods. While the volatility is small in both prior to $2008$ and after $2013$, the leverage effect is quite different in the two periods.  

\begin{figure}[h!]
\includegraphics[width = 140mm, height = 80mm]{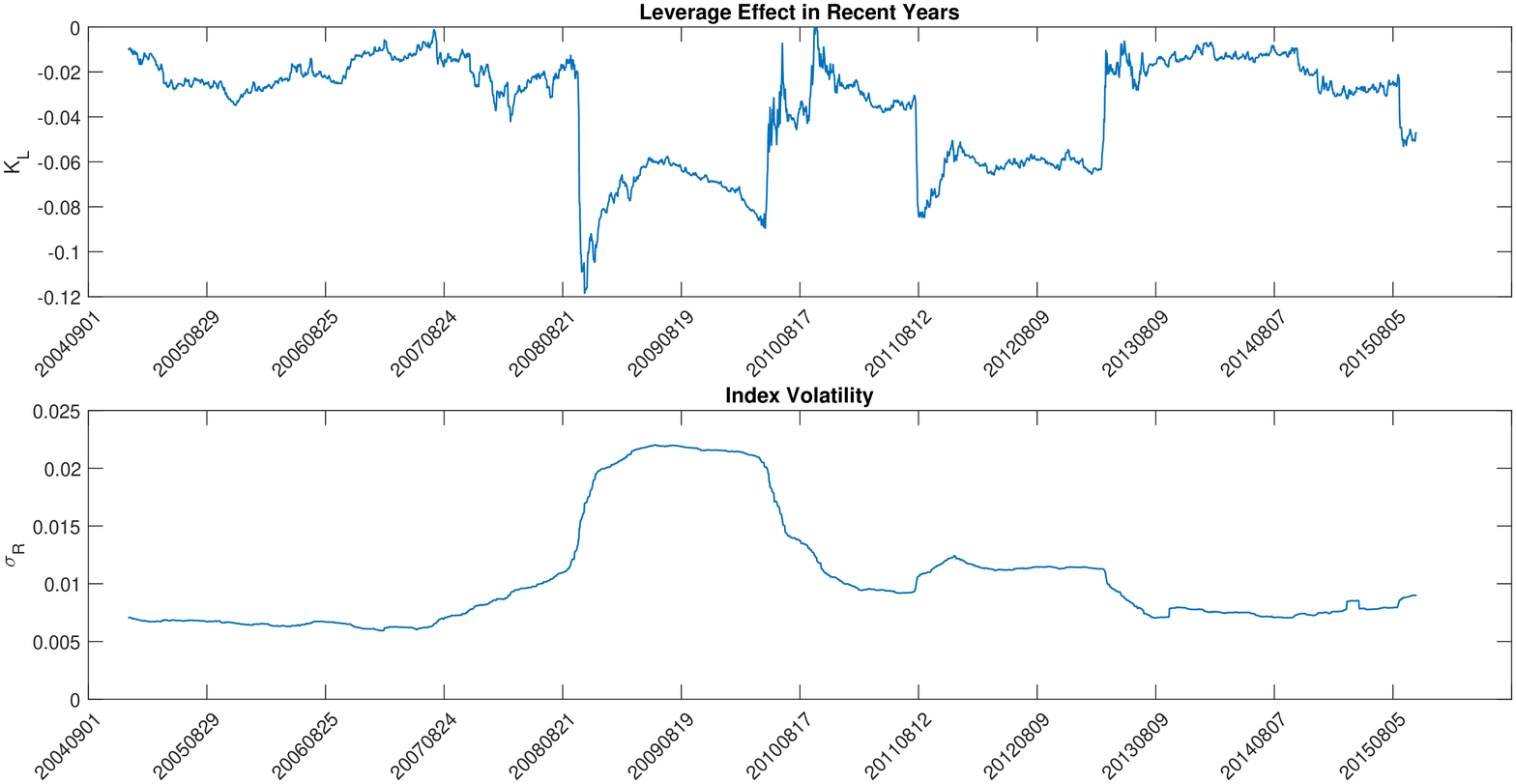}
\caption{ $\bar{K}_L(t)$ and $\bar{\sigma}_R(t)$ change with time }
\end{figure}

\section{Conclusions}
In summary, we propose an alternative framework to model volatility and studied an asymmetric volatility feedback model. This asymmetry is indeed shown in the conditional correlation function $L_{\pm}(\tau)$. We define the feedback functions $K_{\pm}(\tau)$ and compare our first order calculation with the data from Dow Jones components, we find the feedback effect has two time scales. The short one typically within one month is very sensitive to whether the past return is positive or negative, hence produces the leverage effect. The long time scale ranging from 300 days to 1000 days is responsible for the long-term clustered volatility. The feedback is symmetric at this long time scale. We use the moving averaged conditional correlation functions as a market indicator to study the time evolution of those functions. They are not constant and exhibit a rather non-trivial behavior: a pattern of alternative jumps and plateaus. Both the volatility feedback $K_V$ and the leverage feedback $|K_L|$ become large for the market with a large volatility, which may imply the feedback mechanism is not large enough to produce the large deviation observed in the data. We also find the leverage effect becomes stronger after 1987 and a clear pattern switching behavior in recent decade. The model proposed is phenomenological in nature, it will be very interesting to see what market status could result in the features of functions $K_{\pm}(\tau)$ studied here.

%\bibliography{Refs}
\bibliographystyle{unsrt}

\end{document}